\documentclass[aps,prd,preprintnumbers,nofootinbib]{revtex4}
\usepackage{bm}
\usepackage{latexsym}
\usepackage{dcolumn}
\usepackage{amsmath,amsfonts,amssymb}
\usepackage{graphicx,epsfig}
\usepackage{psfrag}
\usepackage{amsthm}
\usepackage[pdftex]{hyperref}
\usepackage[colorinlistoftodos]{todonotes}

\def\be{\begin{eqnarray}}
\def\ee{\end{eqnarray}}

\renewcommand\t{\tilde}

\begin{document}
%opening
%
\title{Towards the
Raychaudhuri Equation Beyond General Relativity
} 
\author{Daniel J Burger}\email{burgerj.daan@gmail.com}
\affiliation{Department of Applied Mathematics and Mathematics,\\
University of Cape Town, \\
Cape Town, South Africa}
\author{Saurya Das}\email{saurya.das@uleth.ca}
\affiliation{Department of Physics and Astronomy,\\
University of Lethbridge, 4401 University Drive,\\
Lethbridge, Alberta T1K 3M4, Canada}
\author{S. Shajidul Haque}\email{shajid.haque@uwindsor.ca}
\affiliation{Department of Physics,\\
University of Windsor,\\
Windsor, Ontario N9A 0C5, Canada}
\author{Nathan Moynihan}\email{nathantmoynihan@gmail.com}
\affiliation{Department of Applied Mathematics and Mathematics,\\
University of Cape Town, \\
Cape Town, South Africa}
\author{Bret Underwood}\email{underwbj@plu.edu}
\affiliation{Department of Physics,\\
Pacific Lutheran University,\\
Tacoma, WA 98447}
\date{\today}

\vspace{-2cm}
\begin{abstract}
In General Relativity, gravity is universally attractive, 
a feature embodied by the Raychaudhuri equation which requires that
the expansion of a congruence of geodesics is always non-increasing,
as long as matter obeys the strong or weak energy conditions.
This behavior of geodesics is an important ingredient in general proofs of singularity theorems,
which show that many spacetimes are singular in the sense of being geodesically incomplete and 
suggest that General Relativity is itself incomplete.
It is possible that alternative theories of gravity, which reduce to General Relativity in some limit,
can resolve these singularities, so it is of interest to consider how the behavior
of geodesics is modified in these frameworks.
We compute the leading corrections to the Raychaudhuri equation for the expansion
due to models in string theory, braneworld gravity, $f(R)$ theories,
and Loop Quantum Cosmology, for cosmological and black hole backgrounds,
and show that while in most cases geodesic convergence is reinforced, in a few cases
terms representing repulsion arise, weakening geodesic convergence and 
thereby the conclusions of the singularity theorems. 
\end{abstract}

\maketitle

\section{Introduction}

General Relativity (GR) is widely expected to be an incomplete theory of dynamical spacetime.
One reason for this is that GR famously predicts its own demise through the existence
of singularities, as demonstrated by the singularity theorems \cite{Penrose:1964wq,Hawking:1969sw}.
An essential physical ingredient of the singularity theorems is that gravity 
is attractive, so that congruences of convergent timelike and null geodesics develop singularities
in finite proper (affine) time.
More specifically, geodesic congruences with timelike $u^M(\tau)$ and null $n^M(\lambda)$ tangent
vector fields are characterized by their expansion $\theta \equiv \nabla_M u^M, \hat \theta \equiv \nabla_M n^M$,
respectively, which satisfy the Raychaudhuri equations \cite{Raychaudhuri}
\be
\frac{d\theta}{d\tau} = -\frac{\theta^2}{D-1}-R_{MN} u^M u^N + ... && \mbox{ (timelike),} \nonumber \\
\frac{d\hat\theta}{d\lambda} = -\frac{\hat \theta^2}{D-2}-R_{MN} n^M n^N + ...  && \mbox{ (null)}
\label{eq:RayGeneral}
\ee
for spacetime dimension $D$, where the additional $+...$ terms are non-positive; see
Appendix \ref{app:RayEq} for details.

For the timelike Raychaudhuri equation, if the so-called ``convergence condition''
\be
R_{MN} u^M u^N \geq 0
\ee
holds, then the expansion of a congruence of geodesics is non-increasing.
Specifically, an initially converging congruence $\theta_i < 0$ develops a singularity $\theta\rightarrow -\infty$
in finite proper time $\tau \sim |\theta_i|^{-1}$. A similar singularity $\hat \theta\rightarrow -\infty$ develops for null convergences
in finite affine time $\lambda \sim |\hat \theta_i|^{-1}$ if the null convergence condition $R_{MN} n^M n^N \geq 0$ is satisfied.
These singularities in the expansion don't necessarily imply a pathology of spacetime themselves; such caustics appear in Minkowski
spacetime, for example \cite{Friedrich:1983vi}.
However, the existence of strictly non-positive contributions to the right-hand side of Eq.(\ref{eq:RayGeneral}),
when combined with other global conditions on the spacetime manifold, form the basis for the general
existence of singularities in cosmological and black hole spacetimes \cite{Penrose:1964wq,Hawking:1969sw}.

Within GR it is possible to convert the convergence conditions into energy conditions on the types of matter.
We can use the (trace-reversed) Einstein equation
\be
R_{MN} = \kappa_D^2 \left(T_{MN} - \frac{1}{D-2} g_{MN} T^M_M\right)
\label{eq:EE}
\ee
(where $\kappa_D^2 \equiv 16\pi G_D$, $G_D$ being the $D$-dimensional Newton's constant)
to substitute in Eq.(\ref{eq:RayGeneral})
\begin{align}
&\frac{d\theta}{d\tau} = -\kappa_D^2 \left(T_{MN} - \frac{1}{D-2} g_{MN} T^M_M \right) u^M u^N + ... & &\mbox{ (timelike), } \nonumber \\
&\frac{d\hat \theta}{d\lambda} = -\kappa_D^2 T_{MN} n^M n^N + ... & &\mbox{ (null) }
\label{eq:RayEq}
\end{align}
where the $+...$ terms are again strictly non-positive
terms.
Thus, all of the terms on the right-hand side of (\ref{eq:RayEq}) are non-positive provided that matter satisfies the respective energy conditions
\be
\left(T_{MN} - \frac{1}{D-2} g_{MN} T^M_M \right) u^M u^N \geq 0 && \mbox{ Strong Energy Condition, } \nonumber \\
T_{MN} n^M n^N \geq 0 && \mbox{ Null Energy Condition.}
\label{eq:EnergyConditions}
\ee
For an isotropic perfect-fluid energy momentum tensor with energy density $\rho$ and pressure $p$, these conditions translate
into
\be
\rho + \frac{D-1}{D-3} p \geq 0 && \mbox{ Strong Energy Condition,} \nonumber \\
\rho + p \geq 0 && \mbox{ Null Energy Condition.}
\label{eq:rhoPEnergyConditions}
\ee
Most known classical matter obeys both the null and strong energy conditions, while vacuum energy $p = -\rho$ violates the strong
energy condition but still saturates the null energy condition.
It may possible to violate the null energy condition with exotic forms of matter \cite{ArkaniHamed:2003uz,Kobayashi:2010cm}, non-minimal coupling \cite{Lee:2007dh,Lee:2010yd},
or quantum gravity effects \cite{Ford:1993bw}, though these approaches often face challenges that we will not explore further here.

However, since we expect corrections to GR of some form, we do not expect the Einstein equations (\ref{eq:EE}) to always hold.
Corrections to Einstein's equations may make it possible to violate the convergence conditions 
$R_{MN} u^M u^N < 0$ and/or $R_{MN} n^M n^N < 0$ without violating
the energy conditions (\ref{eq:EnergyConditions}).
In particular, many corrections to GR appear perturbatively in the form
\be
R_{MN} = \kappa_D^2\left(T_{MN} - \frac{1}{D-2} g_{MN} T^M_M \right) + \lambda H_{MN}\, ,
\label{eq:EECorrection}
\ee
where $\lambda$ controls the strength of the corrections and $H_{MN}$ is a tensor that contains contributions
from the metric, curvature, energy-momentum tensor, or additional fields.
The additional term in Eq.(\ref{eq:EECorrection}) in turn shows up as an additional term on the right-hand side
of the Raychaudhuri equation
\begin{align}
&\frac{d\theta}{d\tau} = -\kappa_D^2 \left(T_{MN} - \frac{1}{D-2} g_{MN} T^M_M \right) u^M u^N + \lambda\ H_{MN} u^M u^N+ ...\, & &\mbox{ (timelike), } \nonumber \\
&\frac{d\hat \theta}{d\lambda} = -\kappa_D^2 T_{MN} n^M n^N + \lambda\ H_{MN}n^M n^N+ ...\, & &\mbox{ (null) }
\label{eq:RayCorrected}
\end{align}
where it may in principle contribute with any sign.
In this paper, we will examine corrections to the Raychaudhuri equations
from four frameworks for corrections to GR: string theory, braneworld gravity, $f(R)$ theories, and
Loop Quantum Cosmology, for cosmological (and in some cases black hole) backgrounds
(see also \cite{Conroy:2016sac,Conroy:2014dja} for a similar analysis in infinite derivative gravity).

While we are motivated by the existence of singularities in these backgrounds, and the promise of these
alternatives to GR for resolving the singularities, we will not attempt to prove the absence of singularities
in this paper. Indeed, a significant amount of work has shown that finding realistic singularity-free 
spacetimes under computational control is quite challenging, and we expect that true singularity resolution
will require physics beyond the perturbative approach of (\ref{eq:EECorrection}).
Instead, we aim for a more modest goal, that of finding corrections to GR that lead to potentially positive
terms on the right-hand side of the Raychaudhuri equation Eq.(\ref{eq:RayGeneral}) for timelike and null
geodesics -- a necessary, but far from sufficient, condition for ultimately resolving spacetime singularities.
%\tred
% The following added in revised version (May 2018)
{If indeed such positive terms are found (as we shall see for a handful of cases), further 
analysis would be required to see, whether:
(a) a generalization of the singularity theorems holds for these cases, or
(b) another criterion for singularity (in lieu of geodesic incompleteness)
can be applied. If none of the above holds, one  
would be forced to conclude that these spacetimes are indeed non-singular. Such an analysis is beyond the scope of this work, however.
}

In Section \ref{sec:StringTheory} we outline a general set of corrections to GR from string theory in $D$ dimensions,
and compute the corrections to the Raychaudhuri equation for black hole and cosmological spacetimes.
In Section \ref{sec:braneworld} we compute the corrections to the 4-dimensional induced Einstein equation for the braneworld scenario
in a cosmological spacetime.
In Section \ref{sec:fR} we compute the corrections to the Einstein equation for so-called $f(R)$ theories, and determine the
form of the corrections in a cosmological background.
In Section \ref{sec:Loop} we compute the corrections to the Raychaudhuri equation in a cosmological background from Loop
Quantum Gravity.
In Section \ref{sec:conclusion} we conclude with some comments on the potential for theories beyond GR to resolve singularities.

\section{String Theory and Gauss-Bonnet Corrections}
\label{sec:StringTheory}

In string theory, higher-order $\alpha'$ corrections induce corrections to the action and, correspondingly, to the Einstein Equations.
In particular, the corrected action at leading order in $\alpha'$ for bosonic, heterotic, type IIA/IIB takes the form \cite{Callan,Zwiebach:1985uq,Gross:1986mw}
\be
S &=& \frac{1}{\kappa_D^2} \int d^Dx \sqrt{-g}\ e^{-2\phi} \left[R_D + 4 (\partial \phi)^2 + \lambda R_{MNPQ}R^{MNPQ} + {\mathcal O}(\alpha'^2)\right] 
\label{eq:RieSq}
\ee
where we are ignoring the antisymmetric rank two tensor $B_{MN}$, $\phi$ is the dilaton, $R_{MNPQ}$ is the $D$-dimensional Riemann tensor, 
%and $\kappa_D^2 = 16\pi G_D$ is the $D$-dimensional gravitational constant, 
and where
\footnote{The leading $\alpha'$ corrections for type II theories occurs first at ${\mathcal O}(\alpha'^3)$ \cite{Becker}.}
\be
\lambda &=& \begin{cases} \frac{1}{2}\alpha' & \mbox{ for bosonic strings} \cr \frac{1}{4}\alpha' & \mbox{ for heterotic strings} \cr 0 & \mbox{ for supersymmetric strings (IIA/IIB).} \cr\end{cases}
\ee
Unfortunately, (\ref{eq:RieSq}) contains higher derivative terms in its equation of motion.
However, it turns out that (\ref{eq:RieSq}) is ambiguous up to field redefinitions of the fields $g_{\mu\nu},\phi$
to next order in $\alpha'$. It is possible to use these field redefinitions to remove the higher order terms in the
equations of motion\footnote{See the discussion in \cite{Zwiebach:1985uq,Gross:1986mw,Turok}.}, giving rise to the action
\be
S_{mod} = \frac{1}{\kappa_D^2} \int d^Dx \sqrt{-g}\ e^{-2\phi} &&\left\{ R_D + 4 (\partial \phi)^2 +\frac{1}{2}\lambda \left[ R_{GB}^2 + 16 \left(R^{MN} - g^{MN} R_D\right) \partial_M \phi \partial_N \phi \right. \right. \nonumber \\
&& \left.\left. -16 \nabla^2 \phi (\partial \phi)^2 + 16 (\partial \phi)^4\right] + {\mathcal O}(\alpha'^2)\right\} + {\mathcal L}_m
\label{eq:StringCorrectAction}
\ee
where we have allowed for additional matter (including potentially a $D$-dimensional cosmological constant) through
${\mathcal L}_m$, and  $R_{GB}^2$ is the 2nd order Gauss-Bonnet combination
\be 
R_{GB}^2 = R_{MNPQ} R^{MNPQ} - 4 R_{MN}R^{MN} + R_D^2 \, .
\label{eq:RGB}
\ee
Setting the dilaton to a constant, we are left with the usual Ricci curvature term and the quadratic Gauss-Bonnet
term in our action
\be
S_{EGB} = \frac{1}{\kappa_D^2}\int d^Dx \sqrt{-g}\left[R_D + \frac{1}{2}\lambda R_{GB}^2\right] + {\mathcal L}_m\, .
\label{eq:EGBAction}
\ee
We will refer to this as the Einstein-Gauss-Bonnet (EGB) gravity action.
The resulting equations of motion are
\be
R_{MN} - \frac{1}{2} g_{MN} R_D = \kappa_D^2 T_{MN} + \frac{1}{2}\lambda \hat H_{MN}\, ,
\label{eq:EinsteinCorrected1}
\ee
where
\be
\hat H_{MN} = && \frac{g_{MN}}{2} R_{GB}^2 - 2 R_D R_{MN} + 4 R_{MA} R^A_N + 4 R^{AB}R_{AMBN}- 2 R_{MABC} R_N^{ABC}\, .
\label{eq:HMNhat}
\ee
In order to put (\ref{eq:EinsteinCorrected1}) in the appropriate form relevant for use the Raychaudhuri equation\footnote{See \cite{Fairoos:2018pee} for an alternative
method of deriving a generalized Raychaudhuri equation for EGB gravity.}, we will
trace-reverse, giving
\be
R_{MN} = \kappa_D^2 \left(T_{MN} - \frac{g_{MN}}{D-2} T^M_M\right) + \frac{\lambda}{2} H_{MN}\, ,
\label{eq:EinsteinCorrected}
\ee
where now
\be
H_{MN} = \frac{g_{MN}}{D-2} R_{GB}^2 - 2 R_D R_{MN} + 4 R_{MA} R^A_N + 4 R^{AB}R_{AMBN}- 2 R_{MABC} R_N^{ABC}\, .
\label{eq:HMN}
\ee

More generally, the terms above are just the leading terms of the more generic Lanczos-Lovelock extensions of gravity \cite{Lanczos:1938sf,Lovelock:1971yv} (see \cite{Padmanabhan:2013xyr} for a review).
The Lanczos-Lovelock extensions include the set of additional terms that can be added to the gravitational action and still
lead to 2nd order equations of motion.
In 4 spacetime dimensions, it can be shown that the Ricci scalar and cosmological constant are the only non-topological
terms that can be added.
In particular, the correction term in (\ref{eq:StringCorrectAction}) is purely topological
in 4-dimensions (it is the Euler characteristic $\chi$ of the spacetime), not contributing to the equation of motion.
For higher dimensions $D>4$, there is a finite series of additional terms, increasing in powers of $R, R_{MN}, R^M_{NPQ}$
(terminating at some order for a given $D$).
We will just focus on this leading order term for now, but will keep in mind that there can be additional corrections to consider.

In order to determine the form of these corrections it is necessary to calculate $H_{MN}$ for specific backgrounds.
For simplicity, we will restrict ourselves to corrections to black hole and cosmological backgrounds; 
in the subsections that follow, we will examine the corrections for these backgrounds in more detail.

\subsection{Black Holes}

The Gauss-Bonnet correction terms (\ref{eq:EinsteinCorrected}) 
appear as part of a pertubative series of corrections, so we expect solutions for some given matter content
to also be described by a pertubative series in $\lambda$: $g_{MN}(\lambda) = g_{MN}^{(0)} + \lambda g_{MN}^{(1)} + ...$, where $g_{MN}^{(0)}$ is the
uncorrected General Relativity solution, $g_{MN}^{(1)}$ is the first-order correction, and so on.
It is remarkable that exact solutions $g_{MN}(\lambda)$ of Einstein-Gauss-Bonnet gravity for black hole backgrounds
are known to all orders in $\lambda$ \cite{Boulware:1985wk,Charmousis:2008kc,Garraffo:2008hu}.
However, since we expect the Gauss-Bonnet corrections to be only the first term in a series of corrections arising from string theory,
we cannot trust these exact solutions beyond ${\mathcal O}(\lambda)$.
One can therefore consider the effects of the Gauss-Bonnet correction terms on the Raychaudhuri equations up to ${\mathcal O}(\lambda)$ 
for black hole backgrounds in two ways:
\begin{enumerate}
\item Evaluate  $R_{MN} u^M u^N \sim \frac{\lambda}{2} H\left[g^{(0)}\right]_{MN} u^M u^N$ on the uncorrected Schwarzschild black hole metric.
%Using solutions to the uncorrected Einstein equations, compute the corrections to the Raychaudhuri equation due to $H_{MN}$, providing a perturbative correction valid up to
%${\mathcal O}(\alpha')$.
\item Evaluate $R[g(\lambda)]_{MN} u^M u^N$ on the known exact metric $g_{MN}(\lambda)$ for black hole backgrounds,
then expand the result to ${\mathcal O}(\lambda)$.
%Using exact solutions to the fully corrected Einstein + Gauss-Bonnet equations themselves (valid to all orders in $\lambda \sim \alpha'$), compute the
%corrections to the Raychaudhuri equation.
\end{enumerate}
(with similar expressions for the null Raychaudhuri equation).
%Since we expect the quadratic Gauss-Bonnet corrections $H_{MN}$ to be only one term
%in a perturbative series of corrections in powers of $\alpha'$, we should
%only trust our corrections to the Raychaudhuri equation up to ${\mathcal O}(\alpha')$ anyway, so the latter approach does not provide any particular advantages.
While both approaches should give identical results, up to ${\mathcal O}(\lambda)$, we will consider and compare both approaches for completeness.

\subsubsection{Perturbative Black Hole Corrections}

Since we are considering actions with a constant dilaton and zero form fields, we will restrict ourselves to
pure gravity solutions.
We thus first begin by considering the perturbative correction terms $H_{MN}$ to a $D$-dimensional black hole solution of the vacuum Einstein equations
\begin{equation}
ds^2 = -f(r) dt^2 + \frac{dr^2}{f(r)} + r^2 d\Omega_{D-2}^2\, ,
\end{equation}
where $d\Omega_{D-2}^2= \hat g_{ij}d\theta^i d\theta^j$ is the metric of a $(D-2)$-dimensional
constant curvature manifold with unit radius (such as a sphere).
The zeroth order black hole solution takes the form
\begin{equation}
f(r) = 1-\frac{\mu}{r^{D-3}}\, ,
\end{equation}
where $\mu$ is related to the mass $M$ of the black hole by
\begin{equation}
M = \frac{(D-2)A_{D-2}}{2\kappa_D^2} \mu
\end{equation}
and $A_{D-2}$ is the area of a unit $(D-2)$ sphere.
Note that for $D=4$ we have $\mu = \kappa_D^2 M/4\pi$.

While the Ricci tensor and scalar vanish $R_{MN} = 0 = R_D$, as is expected for a $D$-dimensional 
Schwarzschild background, the Riemann tensor does not, and gives the only
non-zero contribution to the 
Gauss-Bonnet scalar
\be
R_{GB}^2 = R_{ABCD}R^{ABCD} = (D-3)(D-2)^2(D-1) \frac{\mu^2}{r^{2D-2}}\,,
\ee
and corrections to the Einstein equation,
\be
H_{MN} = \frac{g_{MN}}{D-2} \left[R_{ABCD}R^{ABCD}\right] - 2 R_{MABC}R_N^{ABC}\, .
\ee
We are primarily concerned with the correction terms 
in the $(tt)$ and $(rr)$ directions,
\be
H_{tt} &=&  \frac{f(r)\mu^2}{r^{2D-2}} (D-4)(D-3)(D-2)(D-1)\,; \\
H_{rr} &=& -\frac{1}{f(r)} \frac{\mu^2}{r^{2D-2}} (D-4)(D-3)(D-2)(D-1)\,.
\ee
Note that these correction terms vanish identically $H_{MN}=0$ for
$D \leq 4$, which is what we expect since the Gauss-Bonnet correction is purely topological for $D\leq 4$, and only acts as a dynamical
correction for $D > 4$.

We now consider a radial affine null tangent vector
\be
n^M = \left(\frac{1}{f(r)},\pm 1,\vec{0}\right)\, ,
\ee
where $\pm 1$ corresponds to radially outgoing/ingoing.
The null Raychaudhuri equation takes the form
\be
\frac{d\hat \theta}{d\lambda} = \frac{-\hat \theta^2}{D-2}-|\hat \sigma|^2 - R_{MN}n^M n^N\, .
\ee
Our corrections due to the Gauss-Bonnet term appear on the right hand side, as (recall that $T_{MN} = 0$)
\be
R_{MN}n^M n^N = \frac{\lambda}{2} H_{MN}n^M n^N = \frac{\lambda}{2} H_{tt} n^t n^t + \frac{\lambda}{2} H_{rr} n^r n^r = 0\, .
\ee
Remarkably, the Gauss-Bonnet corrections {\bf vanish} identically everywhere for null rays, implying that the null Raychaudhuri equation for black holes is uncorrected to leading order in $\lambda$.

Finally, consider a timelike geodesic described by the tangent vector
\be
u^M = \left(\frac{1}{f(r)},-\left(\frac{\mu}{r^{D-3}}\right)^{1/2},\vec{0}\right)\,.
\ee
The timelike Raychaudhuri equation takes the form
\be
\frac{d\theta}{d\tau} = -\frac{\theta^2}{D-1} - |\sigma|^2 - R_{MN} u^M u^N\, ,
\ee
where again the corrections to the Raychaudhuri equation due to the Gauss-Bonnet corrections come from the last term
\be
R_{MN} u^M u^N = \frac{\lambda}{2}H_{MN} u^M u^N = \frac{\lambda}{2}H_{tt} u^t u^t + \frac{\lambda}{2}H_{rr} u^r u^r = \frac{\lambda(D-4)(D-3)(D-2)(D-1)}{2} \frac{\mu^2}{r^{2D-2}}\, .
\label{eq:EGBtimelikePert}
\ee
There are a few things to notice about this correction term. First, it does not seem to suffer from any pathologies due to the coordinate singularity of our coordinate system at the horizon; thus, we expect it to be valid throughout the entire spacetime (although the coordinate $r$ will require careful interpretation).
Second, notice that the corrections are {\bf manifestly positive}, thus the corrections make the caustic/conjugate point at
the location of the putative singularity at $r=0$ {\bf worse}, not better!

We have shown that the perturbative EGB corrections to
the Einstein equations do not provide divergence terms
in either the null or timelike Raychaudhuri equations;
in fact, convergence is \emph{strengthened} in the timelike case.
These results above are not particularly surprising 
since it is known that black hole solutions in Einstein-Gauss-Bonnet gravity still posses 
singularities \cite{Boulware:1985wk,Charmousis:2008kc,Garraffo:2008hu},
as we will now examine in more detail.

\subsubsection{Exact Black Hole Solutions}

In the previous section we considered how the Gauss-Bonnet curvature squared terms
lead to perturbative corrections to the Raychaudhuri equation for pure-Einstein gravity black hole 
solutions.
However, exact black hole solutions for Einstein-Gauss-Bonnet gravity are well-known
\cite{Boulware:1985wk,Charmousis:2008kc,Garraffo:2008hu}
(see also \cite{Callan}), so it is also possible to calculate the right-hand-side of the Raychaudhuri
equation for these fully backreacted solutions.

First, let us review the known black hole solutions of $D$-dimensional Einstein-Gauss-Bonnet gravity
\cite{Boulware:1985wk,Charmousis:2008kc,Garraffo:2008hu}.
As before, we will work with the general spherically symmetric metric
\be
ds^2_D = -f(r) dt^2 + \frac{dr^2}{f(r)} + r^2 d\Omega^2_{D-2}\, .
\label{eq:exactBH}
\ee
Solutions to the Einstein-Gauss-Bonnet equations of motion (\ref{eq:EinsteinCorrected})
are \cite{Boulware:1985wk} (see also \cite{Charmousis:2008kc,Garraffo:2008hu})
\be
f(r) = 1 + \frac{r^2}{\hat\lambda} + \sigma \frac{r^2}{\hat\lambda}\sqrt{1+\frac{2\hat\lambda\mu}{r^{D-1}}}\, ,
\label{eq:exactBHf}
\ee
where $\sigma = \pm 1$ labels different branches of solutions, and we have defined
$\hat \lambda = (D-4)(D-3) \lambda$ for convenience.

Consider first the $\sigma = -1$ branch. To lowest order in $\lambda$, the metric is
simply that of a Schwarzschild black hole, and is asymptotically flat as $r\rightarrow \infty$,
reducing to the usual Einstein gravity solution. For this reason, this is usually called the
``Einstein branch.''
In contrast, the $\sigma = +1$ branch,
often called the ``Gauss-Bonnet branch,''
has a non-trivial vacuum structure for $\mu = 0$
\be
f(r)|_{\mu =0} = 1 + \frac{2r^2}{\hat\lambda}\, ,
\ee
corresponding to anti-deSitter space for $\lambda > 0$
with a large negative cosmological constant $\Lambda_{eff} \sim -\hat\lambda^{-1}$
(or deSitter space for $\lambda < 0$).
For non-zero mass, 
the metric in this branch resembles that of Schwarzschild-anti-deSitter space with a negative
mass
\be
f(r) \approx 1 +\frac{\mu}{r^{D-3}} + \frac{2r^2}{\hat\lambda} + {\mathcal O}(\lambda^2)\, .
\ee
The analysis of the stability of this branch requires some care (see
\cite{Boulware:1985wk,Deser:2002jk}).
Note that since the curvature scale of the corresponding
AdS space is non-perturbative in $\lambda$, it is
doubtful that we can trust these solutions as solutions to perturbative string theory.
Further, for $\lambda > 0$ there is a naked
singularity at the origin \cite{Boulware:1985wk}.
For these reasons, we will restrict our analysis to the Einstein branch $\sigma = -1$.

Despite the curvature squared term in EGB gravity,
the Einstein branch solutions (\ref{eq:exactBH},\ref{eq:exactBHf})
still have a curvature singularity at the origin.
This singularity is surrounded by a horizon located
at $r_h$, given by the roots
of the polynomial \cite{Boulware:1985wk}
\be
\hat\lambda r_h^{D-5} + 2 r_h^{D-3} = 2\mu\, .
\ee
For $D > 5$, this horizon always exists; however,
for $D=5$, the existence of the horizon is
guaranteed only for $\mu > \hat\lambda/2$; thus, for
``microscopic'' black holes, even the EGB solutions
posses a naked singularity.
Since the exact EGB black hole solutions still posses a curvature singularity at the origin,
we are not particularly surprised by our perturbative result from
the previous subsection indicating that the corrections to the null and timelike Raychaudhuri equations do not allow for terms that could prevent the formation of conjugate points.

Even though we have the exact EGB solutions, which do possess a curvature singularity, in hand, let us nevertheless compute the contribution of EGB gravity to the right-hand side of the Raychaudhuri equation to explore
the way in which the EGB corrections can affect the formation of conjugate points.

Note that since the solutions (\ref{eq:exactBH},\ref{eq:exactBHf})
are {\bf exact} solutions to the corrected Einstein equations
\be
R_{MN} - \frac{1}{2} g_{MN} R = \kappa_D^2 T_{MN} + \frac{1}{2} \lambda H_{MN}\, ,
\ee
we can compute the curvature term in the null (timelike) Raychaudhuri equation
$R_{MN} N^M N^N$ ($R_{MN} u^M u^N$ respectively) \emph{directly},
without needing to compute the quadratic curvature terms.

For the metric (\ref{eq:exactBH}), we have
\be
R_{tt} &=& \frac{1}{2} f(r) \left(f''(r) + \frac{(D-2)}{r} f(r)\right)\, ; \\
R_{rr} &=& -\frac{1}{2} \frac{1}{f(r)} \left(f''(r) + \frac{(D-2)}{r} f(r)\right)\, ,
\ee
where a prime $'$ denotes a derivative with respect to $r$.
Note that these
vanish for a Schwarzschild solution $f(r) = 1-\frac{\mu}{r^{D-3}}$,
as expected.

A radial affine null vector has the same form as
in the Schwarzschild case
\be
n^M = \left(\frac{1}{f(r)},\pm 1,\vec{0}\right)\, ,
\ee
where now $f(r)$ refers to the EGB corrected form (\ref{eq:exactBHf}).
The curvature term in the Raychaudhuri equation is then
\be
R_{MN} n^M n^N = R_{tt} \left(n^t\right)^2 + R_{rr} \left(n^r\right)^2 = 0\, ,
\ee
which again vanishes identically, as we found previously in the perturbative case.
It is important to note that the vanishing result is independent of the precise functional form
of $f(r)$, and only requires the generic structure of the metric
(\ref{eq:exactBH}).

The geodesic timelike null vector for the metric (\ref{eq:exactBH})
takes the form
\be
u^M = \left(\frac{1}{f(r)},\pm \sqrt{1-f(r)},\vec{0}\right)\, ,
\ee
where we are assuming $f(r) < 1$, which is the case for the Einstein branch of solutions.
The curvature term in the Raychaudhuri equation then takes the form
\be
R_{MN} u^M u^N = R_{tt} \left(u^t\right)^2 + R_{rr} \left(u^r\right)^2 =\frac{1}{2} \left(f''(r) + \frac{(D-2)}{r} f'(r)\right)\, .
\ee
The general expression for $f(r)$ given in (\ref{eq:exactBHf}) is not
particularly illuminating; however, expanding the solution
in powers of $\lambda$, we obtain
\be
R_{MN} u^M u^N \approx \frac{\lambda (D-4)(D-3)(D-2)(D-1) \mu^2}{4r^{2D-2}} + {\mathcal O}(\lambda^2)\, ,
\ee
matching
our perturbative result (\ref{eq:EGBtimelikePert}), as expected.

As in the perturbative analysis, 
we see that the EGB corrections to the
Schwarzschild black hole background give rise to
either vanishing or convergent contributions
to the null and timelike Raychaudhuri equations.
Thus, the EGB corrections themselves, despite being quadratic in curvature, do not alleviate convergence
that leads to conjugate points and singularities in these
spaces.

We have restricted ourself to spherically symmetric black hole solutions
for simplicity; however, since we have not found an improvement in the convergence behavior, we do not expect that deviations from spherical symmetry are likely to produce qualitatively different results.

Black hole backgrounds are not the only spacetimes of interest for the study of singularities and the Raychaudhuri equation. In the next subsection, we will
explore the application of EGB gravity to cosmological
spacetimes.

\subsection{Cosmology}

Let's examine the EGB corrections (\ref{eq:HMN}) for a $D$-dimensional
cosmological background
\footnote{In the Appendix, we consider cosmological models in which spacetime is divided into a $d$-dimensional external spacetime and $m$-dimensional
internal space, each of which with their own time-dependent scale factors.}
\begin{equation}
ds^2 = -dt^2 + a(t)^2 \left(dr^2+ r^2 \hat{d\Omega}_{D-2}^2\right)\, ,
\label{eq:EGBCosmoMetric}
\end{equation}
where $\hat{d\Omega}_{D-2}^2 = \hat g_{ij}d\theta^i d\theta^j$ is the metric of a $(D-2)$-dimensional sphere.
Our matter will consist of a $D$-dimensional perfect fluid
\be
T_{MN} = (\rho+p)u_M u_N + p g_{MN}\, ,
\ee
where $\rho,p$ are the energy density and pressure of the fluid, respectively.

It is straightforward to compute the Gauss-Bonnet scalar
\be
R_{GB}^2 &=& R_{ABCD}R^{ABCD} - 4R_{MN} R^{MN} + (R_D)^2 \nonumber \\
	&=& 4\frac{\ddot a}{a} \frac{\dot a^2}{a^2}(D-3)(D-2)(D-1) + \frac{\dot a^4}{a^4}(D-4)(D-3)(D-2)(D-1)\, ,
\ee
and the correction terms (\ref{eq:HMN})
\be
H_{tt}	&=& 2 \frac{\ddot a}{a}\frac{\dot a^2}{a^2} (D-4)(D-3)(D-1)-(D-4)(D-3)(D-1) \frac{\dot a^4}{a^4}\, ; \\
H_{rr} 	&=&  -2 \frac{\ddot a \dot a^2}{a} (D-4)(D-3) - \frac{\dot a^4}{a^2} (D-4)(D-3)^2\, .
\ee
Notice that these corrections vanish identically for $D \leq 4$, as expected since for lower dimensions the EGB terms are topological and don't contributed
to the equations of motion.

From the radial affine null tangent vector
\be
n^M = \left(\frac{1}{a(t)},\pm \frac{1}{a(t)^2},\vec{0}\right)\, ,
\ee
where again $\pm$ refers to radially outgoing/ingoing rays,
we can compute the corrections to the null Raychaudhuri equation
\be
R_{MN}n^M n^N &=& \kappa_D^2 T_{MN}n^M n^N + \frac{\lambda}{2} H_{MN}n^M n^N \nonumber \\
	&=& \kappa_D^2\frac{(\rho+p)}{a^2}+ \lambda\frac{\ddot a \dot a^2}{a^5} (D-4)(D-3)(D-2)(D-1) - \lambda \frac{\dot a^4}{a^6}(D-4)(D-3)(D-2) \nonumber \\
	&=& \kappa_D^2\frac{(\rho+p)}{a^2}+\lambda \frac{\dot H_D H_D^2}{a^2} (D-4)(D-3)(D-2)(D-1) + \lambda \frac{H_D^4}{a^2} (D-4)(D-3)(D-2)^2\, ,
\label{eq:EGBCosmoNull}
\ee
where we wrote $H_D \equiv \dot a/a$ to simplify our result.

Clearly the first term in (\ref{eq:EGBCosmoNull}) is positive
for matter that obeys the null energy condition, as usual.
However, the second term can be negative if the time variation
of the Hubble parameter $\dot H_D$ is large enough.

In particular, let us consider perturbative solutions for the metric
(\ref{eq:EGBCosmoMetric}) of the form
\be
a(t) \approx a_0(t) + \lambda a_1(t) + ...
\ee
where $a_0(t)$ is the zeroth-order solution to the $\lambda= 0$ Einstein equation
\be
\frac{\dot a_0^2}{a_0^2} &=& \frac{\kappa_D^2}{6} \rho \equiv \left(H_D^{(0)}\right)^2\, ; \\
\frac{\ddot a_0}{a_0} &=& -\frac{\kappa_D^2}{12} (\rho+3p) \equiv \dot H_D^{(0)} + \left(H_D^{(0)}\right)^2\, ,
\ee
and $H_D^{(0)}$ is the corresponding zeroth-order Hubble parameter.
Inserting $H_D^{(0)}$ into (\ref{eq:EGBCosmoNull}),
we obtain the expression
\be
R_{MN} n^M n^N = \kappa_D^2 \frac{(\rho+p)}{a_0^2} -\lambda \kappa_D^4 \frac{(D-4)(D-3)(D-2)}{24 a_0^2}\rho \left(\frac{D+1}{3}\rho+\frac{D-1}p\right) + {\mathcal O}(\lambda^2)\, .
\ee
Assuming an equation of state $p = w\rho$, the second term is
negative for equations of state $w > -\frac{D+1}{3(D-1)}$.
The lower bound is always greater than $-1/2$, so most ordinary matter
will satisfy this condition.
In particular, a $D$-dimensional universe dominated by radiation $w=1/3$
can contribute a divergent term in the null Raychaudhuri equation.

Finally, consider a comoving, proper-time parameterized timelike geodesic described by the tangent vector
\be
u^M = \left(1,0,0,0\right)\, .
\ee
The corrections to the timelike Raychaudhuri equation due to the Gauss-Bonnet corrections come from the last term
\be
R_{MN} u^M u^N &= & \kappa_D^2 \left(T_{MN}u^M u^N+\frac{1}{2} T^M_M\right) + \frac{\lambda}{2} H_{MN} u^M u^N \nonumber \\
	&=& \kappa_D^2 \frac{(D-3)\rho+(D-1)p}{D-2}+\frac{\lambda}{2} (D-4)(D-3)(D-1) \left[2\dot H_D + H_D^2\right]H_D^2\, ,
\ee
where again we substituted $H_D \equiv \dot a/a$ to simplify our result.
The first term is the typical term for matter, and
is positive for matter that obeys the strong energy condition.
The last term, proportional to $\lambda$, is the new contribution;
we see that for $\dot H_D \sim -H_D^2$ (as is the case for any $a(t) \sim t^n$ time-dependence), 
this term can be negative, giving rise
to a \emph{positive} divergent term in the timelike Raychaudhuri equation,
potentially opposing convergence.

We have seen in both the null and timelike Raychaudhuri equations
the existence of terms that can give rise to a positive (divergent)
contribution to the divergence.
This does not necessarily mean that cosmology with EGB gravity can evade the singularity problem, 
just that the usual singularity theorems do
not apply in a straightforward way to these backgrounds.

In particular, consider the EGB-corrected equations of motion
for the metric
(\ref{eq:EGBCosmoMetric}) from (\ref{eq:EinsteinCorrected}).
Examining the $(tt)$ component of (\ref{eq:EinsteinCorrected}) we have
\be
\frac{1}{2} (D-2)(D-1) \frac{\dot a^2}{a^2} = \kappa_D^2 \rho - \frac{\lambda}{4} (D-4)(D-3)(D-2)(D-1) \frac{\dot a^4}{a^4}\, .
\label{eq:EGBFriedmann}
\ee
For $\lambda\kappa_D^2 \rho \ll 1$, solutions to (\ref{eq:EGBFriedmann})
take the usual Einstein form; 
however, for $\lambda \kappa_D^2 \rho \gg 1$, as
we would expect to occur in the early universe near a big bang singularity,
we have instead
\be
\frac{\dot a^2}{a^2} \sim \sqrt{\frac{\kappa_D^2\rho}{\lambda}}\,.
\ee
It is clear that a curvature singularity still exists where $a(t)\rightarrow 0,
R_D \rightarrow \infty$, despite the presence of the higher
curvature terms. \\

%%%%%%%%%%%%%%%%%%%%%%%%%%%%%%%%%SHAJID%%%%%%%%%
%%%%%%%%%%%%%%%%%%%%%%%%%%%%%%%%%%%%%%%%%%%%%%%%
\section{Braneworld Gravity}
\label{sec:braneworld}
We will consider a 5-dimensional bulk spacetime (with 5-dimensional cosmological constant $\Lambda_5$) 
with coordinates $(t,\vec{x},y)$ and metric $g_{MN}$ given by \cite{Brax:2004xh}
\be
ds^2 = a(t,y)^2 b(t,y)^2 \left(-dt^2 + dy^2\right) + a^2(t,y)(dr^2 + r^2 d\Omega^2)\, ,
\label{eq:BraneMetric}
\ee
which is sufficiently general to capture the backreaction of the singular brane on the bulk
as well as 4-dimensional homogeneous and isotropic cosmological evolution.
The extra dimension has the range $y \in (-\infty,\infty)$; however, we will additionally impose a $\mathbb Z_2$ symmetry \cite {Horava:1995qa}
$y \rightarrow -y$, so that the covering space is reduced to $y \in [0,\infty)$, as in the RS I scenario \cite{Randall:1999ee}. 
We embed a (singular) 3-brane at $y=0$ with unit normal $n^M = \left(0,0,0,0, \frac{1}{ab}\right)$
and induced metric $q_{\mu\nu} = g_{\mu\nu} - n_\mu n_\nu$; we will use capital Latin letters $M,N$ for 5-dimensional
coordinate indicies and lowercase Greek letters $\mu,\nu$ for 4-dimensional brane indicies.
The 5-dimensional energy-momentum tensor
has the form
\be
^{(5)}T_{MN} = -\Lambda_5 g_{MN} + \delta^\mu_M \delta^\nu_N S_{\mu\nu} \delta(y), \hspace{.2in} S_{\mu\nu} = -\sigma q_{\mu\nu}  + \tau_{\mu\nu}
\label{eq:BraneBulkEMTensor}
\ee
where $\sigma$ is the brane tension
and $\tau_{\mu\nu}$ is the brane matter energy momentum tensor.
Note that the usual Minkowski RS I model has the solution
\be
ds^2 = e^{-2K(z)} \eta_{\mu\nu} dx^\mu dx^\nu + dz^2\, .
\label{eq:RSmetric}
\ee
This can be obtained from (\ref{eq:BraneMetric}) under the limits and identifications
\be
b(t,y) &\rightarrow & 1\, ; \nonumber \\
\label{eq:RSconditions} a(t,y) &\rightarrow & a(y) = e^{-K(z)}\, ; \\
z &=& \int_0^\infty a(y) dy\, . \nonumber
\ee

Gravity on the 3-brane hypersurface is induced by its embedding in the extra dimensions; in particular,
the induced 4-dimensional Einstein equations on the brane are \cite{Shiromizu:1999wj}
\be
^{(4)}R_{\mu\nu} - \frac{1}{2} q_{\mu\nu}\ ^{(4)}R = - \Lambda_4 q_{\mu\nu} + \kappa_4^2 \tau_{\mu\nu} + \kappa_5^4 \pi_{\mu\nu} - E_{\mu\nu},
\label{eq:inducedEE}
\ee
where $\Lambda_4= \frac{1}{2} \kappa_5^2 \left ( \Lambda_5 + \frac{1}{6} \kappa_5^2  \sigma^2 \right)$ and $\kappa_4^2= \frac{ \kappa_5^4 \sigma}{6}$,
and the new tensors $\pi_{\mu\nu}, E_{\mu\nu}$ are defined as
\be
\pi_{\mu \nu}&=& \lim_{y\rightarrow 0}\left[-\frac{1}{4} \tau_{\mu \alpha} \tau^{\alpha}_{\nu} +\frac{1}{12} \tau \tau_{\mu \nu} +\frac{1}{8} q_{\mu \nu} \tau_{\alpha \beta} \tau^{\alpha \beta}-\frac{1}{24} q_{\mu \nu} \tau^2\right]\, ; \\
E_{\mu\nu} &=& \lim_{y\rightarrow 0} \left[\ ^{(5)}C_{\mu\alpha\nu\beta} n^\alpha n^\beta\right] ,
\ee
where $^{(5)}C_{MNPQ}$ is the 5-dimensional Weyl tensor; see Appendix \ref{sec:Weyl} for more details.
The first two terms on the right-hand side of (\ref{eq:inducedEE}) are just the usual 4-dimensional cosmological constant
and energy-momentum tensor sources, while the last two terms arise from the extra dimensional embedding of the brane.
In order to put this into a form suitable for use with the Raychaudhuri equation we trace-reverse (\ref{eq:inducedEE})
\be
^{(4)}R_{\mu\nu} &=& \Lambda_4 q_{\mu\nu} + 8\pi G_N \left(\tau_{\mu\nu} - \frac{1}{2}q_{\mu\nu} \tau^\mu_\mu\right)
	+ \kappa_5^4 \left(\pi_{\mu\nu} - \frac{1}{2} q_{\mu\nu} \pi^\mu_\mu\right) - \left(E_{\mu\nu} - \frac{1}{2}q_{\mu\nu} E^\mu_\mu\right) \\
    &=& \Lambda_4 q_{\mu\nu} + 8\pi G_N \left(\tau_{\mu\nu} - \frac{1}{2}q_{\mu\nu} \tau^\mu_\mu\right) + H_{\mu\nu}\, ,
\label{eq:inducedRmunu}
\ee
where we rewrote the last two terms as a correction term $H_{\mu\nu}$ so that the corrections take the form (\ref{eq:EECorrection}) 
as outlined in the Introduction.
The timelike and null 4-dimensional Raychaudhuri equations thus have the new terms arising from the braneworld
\be
\label{eq:BraneRayTime}\frac{d\theta}{d\tau} &=& - ^{(4)}R_{\mu\nu} u^\mu u^\nu +...=  - \kappa_5^4 \left(\pi_{\mu\nu} u^\mu u^\nu+\frac{1}{2}\pi^\mu_\mu\right) + \left(E_{\mu\nu} u^\mu u^\nu+\frac{1}{2} E^\mu_\mu\right) + ...\, ;\\
\label{eq:BraneRayNull}\frac{d\hat\theta}{d\lambda} &=& - ^{(4)}R_{\mu\nu} n^\mu n^\nu + ...= - \kappa_5^4 \pi_{\mu\nu} n^\mu n^\nu + E_{\mu\nu} n^\mu n^\nu+ ...\, ,
\ee
where $...$ on the right-hand sides denote non-positive terms.

Assuming the brane energy-momentum tensor is of the perfect fluid form, with energy density $\rho(t)$ and pressure $p(t)$,
we can immediately compute the $\pi_{\mu\nu}$ tensor contribution for the metric (\ref{eq:BraneMetric})
\begin{equation}
\begin{array}{ll}
\pi_{tt} = \frac{1}{12} \rho^2(t)\ a^2(t,0)\ b^2(t,0); & \pi_{rr} = \frac{1}{12} \rho(t) \left(2p(t)+\rho(t)\right) a^2(t,0); \cr
\pi_{\theta\theta} = \frac{1}{12} \rho(t) \left(2p(t)+\rho(t)\right) a(t,0)^2 r^2;\hspace{.2in}\ & \pi_{\phi\phi} = \frac{1}{12} \rho(t) (2p(t)+\rho(t)) ;a^2(t,0) r^2 \sin^2\theta\, ,\cr
\end{array}
\end{equation}
where we evaluated the terms at $y = 0$, while we compute $E_{\mu\nu}$ directly from the  metric (\ref{eq:BraneMetric})
\be
E_{tt}& = & \lim_{y\rightarrow 0} \frac{1}{2}\left[\frac{b''}{b} - \frac{b'^2}{b^2} - \frac{\ddot b}{b} + \frac{\dot b^2}{b^2}\right] \equiv \lim_{y\rightarrow 0} \frac{1}{2} \beta(t,y); \\
E_{rr}& = & \frac{1}{6}\lim_{y\rightarrow 0} \beta(t,y) = \frac{E_{\theta\theta}}{r^2} = \frac{E_{\phi\phi}}{r^2 \sin^2\theta}\, ,
\ee
where a dot denotes a derivative with respect to $t$, and a prime denotes a derivative with respect to $y$.
All of the components of the $E_{\mu\nu}$ tensor are proportional to the quantity $\beta(t,y)$, involving a second derivative $b''(t,y)$.

In order to evaluate the common quantity $\beta(t,y)$ appearing in $E_{\mu\nu}$ in the limit $y\rightarrow 0$, we need to use 
the 5-dimensional Einstein equations $^{(5)}R_{MN} - \frac{1}{2} g_{MN} ^{(5)}R = \kappa_5^2\ ^{(5)} T_{MN}$ for the
metric (\ref{eq:BraneMetric}) and sources (\ref{eq:BraneBulkEMTensor}), which take the form \cite{Brax:2004xh}
\be
\label{eq:5dEEtt}(tt) && 3 \left[2\frac{\dot a^2}{a^2} + \frac{\dot a \dot b}{ab} - \frac{a''}{a} + \frac{a'b'}{ab}\right] = a^2 b^2 \kappa_5^2\left[\Lambda_5 + (\rho+\sigma) \delta(y)\right]\, ; \\
\label{eq:5dEEyy}(yy) && 3 \left[\frac{\ddot a}{a} - \frac{\dot a \dot b}{ab} - 2\frac{a'^2}{a^2} - \frac{a'b'}{ab}\right] = a^2 b^2\kappa_5^2 \Lambda_5\, ; \\
\label{eq:5dEEty}(ty) && 3 \left[-\frac{\dot a'}{a} + 2 \frac{\dot a a'}{a^2} + \frac{\dot a b'}{ab} + \frac{a'\dot b}{ab}\right] = 0\, ; \\
\label{eq:5dEEij}(rr)\ \&\ (ij) && \left[3\frac{\ddot a}{a} + \frac{\ddot b}{b} - \frac{\dot b^2}{b^2} - 3 \frac{a''}{a} - \frac{b''}{b} + \frac{b'^2}{b^2}\right] = a^2 b^2\kappa_5^2\left[\Lambda_5 + (\sigma- p) \delta(y)\right]\, .
\ee
Integrating the $(tt)$ and $(ij)$ equations across the brane gives rise to
\be
\left.\frac{a'}{a}\right|_{y=0} = \frac{1}{6} a(t,0) b(t,0)\ \kappa_5^2(\rho+\sigma), \hspace{.4in} \left. \frac{b'}{b} \right|_{y=0} = -\frac{1}{2} a(t,0) b(t,0)\, \kappa_5^2(\rho+p)\, .
\label{eq:primeRestrict}
\ee
In addition, the restriction of the $(ty)$ component of the Einstein equations to the brane at $y=0$
takes the form of the usual energy-momentum conservation equation on the brane
\be
\dot \rho + 3 \left.\frac{\dot a}{a}\right|_{y=0} (\rho+p) = 0\, .
\label{eq:EMConservation}
\ee
Finally, restricting the $(yy)$ component of the Einstein equations to the brane at $y=0$ 
we can obtain a
``Hubble''-like equation
\be
\left(\frac{1}{ab} \frac{\dot a}{a}\right)^2 = \kappa_5^4 \frac{(\rho+\sigma)^2}{36} + \kappa_5^2\frac{\Lambda_5}{6} + \frac{\mu}{a^4}\, ,
\label{eq:Hubble}
\ee
Where the last term arises as an integration constant and is known as the ``dark radiation'' term \cite{Mukohyama:1999qx,Ida:1999ui,Ichiki:2002eh}.
We will assume $\mu \geq 0$.

We can rearrange (\ref{eq:5dEEij}) to write
\be
\beta(t,y) = \frac{b''}{b} - \frac{b'^2}{b^2} - \frac{\ddot b}{b} + \frac{\dot b^2}{b^2} = 3 \left(\frac{\ddot a}{a} - \frac{a''}{a}\right) - a^2 b^2 \kappa_5^2(\Lambda_5 -  p \delta(y))\, .
\ee
Subtracting (\ref{eq:5dEEyy}) and (\ref{eq:5dEEtt}), this then becomes
\be
\beta(t,y)= 6 \left(\frac{\dot a^2}{a^2} - \frac{a'^2}{a^2}\right) - a^2 b^2 \kappa_5^2\left(\Lambda_5 + (\rho+p)\delta(y)\right)\,.
\ee
Using the jump condition (\ref{eq:primeRestrict}) and the Hubble equation (\ref{eq:Hubble}), this simplifies considerably to
\be
\beta(t,y) = 6 \mu \frac{b^2}{a^2}- a^2 b^2 \kappa_5^2 (\rho+p) \delta(y)\, .
\ee
The $E_{\mu\nu}$ tensor is evaluated in the limit $y\rightarrow 0$; since the delta-function does not contribute in this 
limit, we have
\be
E_{\mu\nu} \propto \lim_{y\rightarrow 0} \beta(t,y) = 6\mu \frac{b^2}{a^2}\, ,
\ee
thus the Weyl tensor term $E_{\mu\nu}$ only potentially contributes to the induced Einstein equations on the brane through
the dark radiation term. However, the induced
Einstein equations (\ref{eq:inducedEE}) are not closed,
and must be supplemented by
additional constraints \cite{Shiromizu:1999wj}.
Examining these constraints for our background
and matter content in Appendix \ref{sec:Weyl}, we find that the dark radiation term
$\mu$ must vanish, so that the $E_{\mu\nu}$ contributions
vanish as well.

In order to evaluate the correction terms (\ref{eq:BraneRayTime},\ref{eq:BraneRayNull}), we will take the
proper-time parameterized timelike vector and affine null vectors on the brane to be
\begin{eqnarray}
u^{\mu} &=& \left( \frac{1}{a(t,0)\, b(t,0)}, 0,0,0\right)\, , \\
n^{\mu} &=& \left(\frac{1}{a(t,0)\, b(t,0)}, \pm\frac{1}{a(t,0)}, 0, 0  \right)\, ,
\end{eqnarray}
where the $\pm$ refers to radially outgoing/ingoing null rays.
We can then compute the additional braneworld terms
\be
H_{\mu\nu} u^\mu u^\nu &=& \kappa_5^4 \rho \left(\rho + \frac{3}{2} p\right); \\
H_{\mu\nu} n^\mu n^\nu &=& \kappa_5^4 \frac{\rho^2}{12}\, .
\ee
We see that the braneworld terms only contribute positively to the additional terms for null vectors, while
the additional contributions for timelike vectors can be negative if $p < -2/3\, \rho$.
Including these results, we have the following expressions for the timelike and null Raychaudhuri equations on the brane
\begin{eqnarray}
\label{eq:BraneRayTime2}\frac{d\theta}{d\tau} &=& \Lambda_4-\frac{1}{2} G_N\left(\rho+3p\right) - \kappa_5^4\rho\left[\rho+\frac{3}{2}p\right]  + ... \, ; \\
\frac{d\hat\theta}{d\lambda} &=& -G_N(\rho+p)-\kappa_5^4\frac{\rho^2}{12} + ... \, ,
\end{eqnarray}
where again $...$ refer to strictly non-positive terms.
Thus, we see that the corrections to the induced Einstein equations on the brane due to the braneworld embedding of the brane only contribute
to convergence for the null Raychaudhuri equation. However, the additional contributions do add divergent (positive) terms to the right-hand
side of the timelike Raychaudhuri equation; we note, though, that the second term in Eq.(\ref{eq:BraneRayTime2}), which is present in un-modified
4-dimensional General Relativity, is also divergent in the presence of matter that violates the strong energy condition, $p < -1/3\ \rho$.

%[This is in general agreement with specific cosmological solutions of the induced Einstein equations, which still
%possess a Big Bang cosmological singularity.]
Is the presence of the additional positive term in Eq.(\ref{eq:BraneRayTime2}) arising from induced gravity on the brane sufficient to avoid a singularity?
Rewriting the Friedmann-like equation Eq.(\ref{eq:Hubble}) using the relations
$\Lambda_4= \frac{1}{2} \kappa_5^2 \left ( \Lambda_5 + \frac{1}{6} \kappa_5^2  \sigma^2 \right)$ and $\kappa_4^2= \frac{ \kappa_5^4 \sigma}{6}$
and by reparameterizing the time variable on the brane into cosmic time, $dt' = a(t,0) b(t,0)\ dt$, we have \cite{Binetruy:1999hy, Binetruy:1999ut}
\be
\left(\frac{da/dt'}{a}\right)^2 = \frac{\kappa_4^2}{3} \rho \left( 1+ \frac{\rho}{2 \sigma} \right) + \frac{\Lambda_4}{3} \, . 
\label{eq:braneFriedmann}
\ee
Notice that at late times, when the energy density is much smaller than the brane tension $\rho \ll \sigma$, we recover
the usual Friedmann equation of General Relativity.
However, at early times $\rho \geq \sigma$, assuming an equation of state $\rho = wp$ ($w\neq 1$) implies $\rho \propto a^{-3(1+w)}$ (recall that local energy
conservation on the brane Eq.(\ref{eq:EMConservation}) is unchanged)
so that \citep{Langlois:2004kc,Barrow:2001pi}:
\be
\frac{da/dt'}{a} \sim \frac{a^{-3(1+w)}}{\sqrt{\sigma}}\, ; \hspace{.4in} \Rightarrow \hspace{.4in} a \sim \left((1+w)\frac{t'}{\sqrt{\sigma}}\right)^{\frac{1}{3(1+w)}} \, ,
\ee
%which implies that the scale factor:
%\be
%a \sim \left((1+w)\frac{t}{\sqrt{\sigma}}\right)^{\frac{1}{3(1+w)}}
%\ee
which diverges as $t'\longrightarrow 0$.
Thus, we expect the Big Bang singularity persists in braneworld induced gravity models as well.

\section{f(R) Theories}
\label{sec:fR}

Higher curvature gravity theories, collectively known as $f(R)$ theories, have been studied both for its improved short distance properties,
although at the expense of introducing ghosts \cite{Stelle:1976gc}, 
as well as for their potential for avoiding singularities, and as 
a model of inflation and an alternative model for dark energy. We start with the action for $f(R)$ gravity in the so-called Jordan frame \cite{Sotiriou:2008rp,Albareti:2012va,Santos:2016vjg,Nojiri:2017ncd}
\begin{eqnarray}
S = \frac{1}{\kappa_4^2} \int d^4 x \sqrt{-g}~f(R) + S (g_{\mu\nu}, \psi) ,
\end{eqnarray}
where the last term represents
matter action.

From this action, variation of the metric gives rise to the
modified Einstein equations \cite{Albareti:2012va}
\be
R_{\mu\nu}\ f'(R) = \kappa_4^2 T_{\mu\nu}+\frac{1}{2} g_{\mu\nu} f(R) + \left(\nabla_\mu \nabla_\nu - g_{\mu\nu} \Box \right) f'(R)\, .
\ee
From this we can write the null and timelike Raychaudhuri equations for $f(R)$ gravity:
\begin{eqnarray}
\label{fRRayNull1}
\frac{d\hat\theta}{d\lambda} &=& -\frac{\hat\theta^2}{2} - \left[\frac{\kappa_4^2}{f'(R)} T_{\mu\nu} + \frac{1}{2}g_{\mu\nu} \frac{f(R)}{f'(R)} + \frac{1}{f'(R)} \left(\nabla_\mu \nabla_\nu - g_{\mu\nu} \Box\right)f'(R)\right] n^\mu n^\nu\, ; \\
\frac{d\theta}{d\tau} &=& -\frac{\theta^2}{3} - \left[\frac{\kappa_4^2}{f'(R)} T_{\mu\nu} + \frac{1}{2}g_{\mu\nu} \frac{f(R)}{f'(R)} + \frac{1}{f'(R)} \left(\nabla_\mu \nabla_\nu - g_{\mu\nu} \Box\right)f'(R)\right] u^\mu u^\nu\, ,
\label{fRRay1}
\end{eqnarray}
in terms of the expansions $\hat \theta, \theta$ 
for null and timelike geodesics $n^\mu(\lambda),u^\mu(\tau)$ parameterized
by affine and proper times $\lambda,\tau$, respectively.

Since we are interested in singularities with larger curvatures, we
will consider an $f(R)$ theory consisting of the usual scalar curvature
plus a perturbative
correction of higher power in the curvature 
 $f(R)=R [ 1 + (\ell_{Pl}^2 R)^{n-1}]$, where we will take the Planck
 length $\ell_{Pl}$ to control the strength of the corrections, and
$n$ is a positive integer. 
Since we are interested in the large curvature regime,
i.e. $R\gg 1/\ell_{Pl}^2$, we will consider the case where
the second term in $f(R)$ dominates so that $f \sim R^n$.
In this case, the equations of motion are dominated by the higher curvature
correction terms, which act as an effective energy-momentum tensor.

%\noindent {\bf BU: Resume previous work. Note that I inserted factors of $\ell_{Pl}$ as necessary.}

Substituting this into the Raychaudhuri equations (\ref{fRRayNull1},\ref{fRRay1}),
we obtain
\begin{eqnarray}
&& \frac{d\hat\theta}{d\lambda} = -\frac{\hat\theta^2}{2} - 
%\frac{1}{\phi} 
\left[
 \frac{\kappa_4^2 T_{\mu\nu}}{2 n\, \ell_{Pl}^{2n-2} R^{n-1}} 
+ \frac{R}{2n}~g_{\mu\nu}
%- \frac{n-2}{2} \left( \frac{\phi}{n} \right)^{1/(n-1)}  
+ \frac{1}{n\, R^{n-1}} \left( \nabla_\mu\nabla_\nu  - g_{\mu\nu}\Box \right) R^{n-1}
\right]n^\mu n^\nu~.;
\label{frRENull} \\
&& \frac{d\theta}{d\tau} = -\frac{\theta^2}{3} - 
%\frac{1}{\phi} 
\left[
 \frac{\kappa_4^2 T_{\mu\nu}}{2 n\, \ell_{Pl}^{2n-2} R^{n-1}} 
+ \frac{R}{2n}~g_{\mu\nu}
%- \frac{n-2}{2} \left( \frac{\phi}{n} \right)^{1/(n-1)}  
+ \frac{1}{n\, R^{n-1}} \left( \nabla_\mu\nabla_\nu  - g_{\mu\nu}\Box \right) R^{n-1}
\right]u^\mu u^\nu~. 
\label{frRE} %\\
%
%&& =-\frac{\theta^2}{3} - \frac{1}{nR^{n-1}} \left[{\kappa} T_{\mu\nu} - \frac{n-2}{2} R ~g_{\mu\nu}
%+ n \left( \nabla_\mu\nabla_\nu - g_{\mu\nu}\Box \right) R^{n-1} \right]u^\mu u^\nu
\end{eqnarray}
Note that (i) the sign of the second term  in square brackets
in each equation 
is opposite to the $T_{\mu\nu}$ term, and (ii) the sign of the last term 
in each equation depends on the spacetime under consideration. 

Next, we consider homogeneous and isotropic cosmological backgrounds with the metric
\begin{eqnarray}
ds^2 = -dt^2 + a(t)^2 \left[ \frac{dr^2}{1-kr^2} + r^2 \left( d\theta^2 + \sin^2\theta d\phi^2 \right) \right]\, ,
\label{frw2}
\end{eqnarray}
and corresponding scalar curvature
\begin{eqnarray}
&& R = 6\left( \frac{\ddot a}{a}  + \frac{\dot a^2}{a^2} + \frac{k}{a^2} \right) \, .
\end{eqnarray}
The null Raychaudhuri equation (\ref{frRENull}), using $n^\mu n^\nu g_{\mu\nu} = 0$, becomes
\begin{eqnarray}
\frac{d\hat\theta}{d\lambda} = -\frac{\hat \theta^2}{2} - \frac{\kappa_4^2 T_{\mu\nu}n^\mu n^\nu}{2n\ \ell_{Pl}^{2n-2} R^{n-1}}+ \frac{1}{n R^{n-1}} \left(n^{t}\right)^2 \partial_t^2 R^{n-1}\, ,
\end{eqnarray}
while the timelike Raychaudhuri equation (\ref{frRE}) becomes
\begin{eqnarray}
&&\frac{d\theta}{d\tau} =
-\frac{\theta^2}{3} - 
\left[
\frac{\kappa_4^2\ T_{\mu\nu}}{2 n\, \ell_{Pl}^{2n-2} R^{n-1}} 
+ \frac{3}{n} 
\left( \frac{\ddot a}{a}  + \frac{\dot a^2}{a^2} + \frac{k}{a^2} \right)
~g_{\mu\nu}
\right]u^\mu u^\nu\, .
\end{eqnarray}
When the cosmological evolution is dominated by the curvature corrections,
we have an effective perfect fluid
\begin{eqnarray}
&& p_{\rm eff}= w_{\rm eff} \rho_{\rm eff},~~w_{\rm eff} = - \frac{6n^2-7n-1}{6n^2-9n+3}\,, \label{weff}
\end{eqnarray}
with scale factor ($w_{\rm eff} \neq 1$):
\begin{eqnarray}
&& 
a = a_0 t^{\frac{2}{3(1+w_{\rm eff})}},
\frac{\dot a}{a} = \frac{2}{3(1+w_{\rm eff})t},
~\frac{\ddot a}{a}= - \frac{2(1+3w_{\rm eff})}{9(1+w_{\rm eff})^2t^2},~
\frac{\ddot a}{a}  + \frac{\dot a^2}{a^2} + \frac{k}{a^2}  
=\frac{2(1-3w_{\rm eff})}{9(1+w_{\rm eff})^2t^2} + \frac{k}{a_0^2 t^{4/3(1+w_{\rm eff})}} .~
\end{eqnarray}
We can substitute this solution for the scale factor into the Raychaudhuri equations to obtain
\begin{eqnarray}\label{rew2}
\frac{d\hat\theta}{d\lambda} && = -\frac{\hat\theta^2}{2} + \frac{1}{n\ \left\{\frac{2(1-3w_{\rm eff})}{3n(1+w_{\rm eff})^2t^2} + \frac{3k}{n~a_0^2~t^{4/3(1+w_{\rm eff})}}  \right\}^{n-1}}\times \\ &&\left[-\frac{\kappa_4^2 T_{\mu\nu} n^\mu n^\nu}{2\ell_{Pl}^{2n-2} } + \left(n^t\right)^2 \partial_t^2 \left\{\frac{2(1-3w_{\rm eff})}{3n(1+w_{\rm eff})^2t^2} + \frac{3k}{n~a_0^2~t^{4/3(1+w_{\rm eff})}}  \right\}^{n-1}
\right]\,;  \nonumber \\
\frac{d\theta}{d\tau} &&=
-\frac{\theta^2}{3} - 
\frac{{\kappa_4^2}~\rho }{2n\, \ell_{Pl}^{2n-2} \left\{\frac{2(1-3w_{\rm eff})}{3n(1+w_{\rm eff})^2t^2} + \frac{3k}{n~a_0^2~t^{4/3(1+w_{\rm eff})}}  \right\}^{n-1}} 
+  \left\{\frac{2(1-3w_{\rm eff})}{3n(1+w_{\rm eff})^2t^2} + \frac{3k}{n~a_0^2~t^{4/3(1+w_{\rm eff})}}  \right\}\,, 
\label{rew3}~ %\\
%
%&& p=-\rho,~a = a_0 e^{Ht},~\frac{\dot a}{a} = H,~\frac{\ddot a}{a}= H^2,
%\frac{\ddot a}{a}  + \frac{\dot a^2}{a^2} + \frac{k}{a^2} = 2H^2 + \frac{k e^{-2Ht}}{a_0^2} ,~
%w=-1,~H={\mbox constant} \label{rew1} \\
%%
%&&\dot\theta =-\frac{\theta^2}{3} - 
%\left[\frac{{\kappa} T_{\mu\nu}}{ n \left\{ 12H^2 + \frac{6 k e^{-2Ht}}{a_0^2}  \right\}^{n-1}}+ \frac{3}{n} \left\{
%2 H^2 + \frac{k e^{-2Ht}}{a_0^2}\right\}~g_{\mu\nu}
%\right]u^\mu u^\nu,~w =- 1,~H={\mbox constant}~.\label{rew2}
\end{eqnarray}
where we considered the rest frame of the fluid
streamlines, for which $T_{\mu\nu} u^\mu u^\nu =\rho$, and used
$u^2=-1$.

%Therefore in the rest frame of the fluid , we get from Eqs.(\ref{rew0}):
%
%\begin{eqnarray}
%&&\frac{d\theta}{d\tau} =
%-\frac{\theta^2}{3} - 
%\frac{{\kappa_4^2}~\rho }{2n\, \ell_{Pl}^{2n-2} \left\{\frac{2(1-3w_{\rm eff})}{3n(1+w_{\rm eff})^2t^2} + \frac{3k}{n~a_0^2~t^{4/3(1+w_{\rm eff})}}  \right\}^{n-1}} 
%+  \left\{\frac{2(1-3w_{\rm eff})}{3n(1+w_{\rm eff})^2t^2} + \frac{3k}{n~a_0^2~t^{4/3(1+w_{\rm eff})}}  \right\}\,. 
%\label{rew3}
%
%&&\dot\theta =
%-\frac{\theta^2}{3} - \left[
%\frac{{\kappa} \rho }{n \left\{  12 H^2 + \frac{6ke^{-2Ht}}{a_0^2}  \right\}^{n-1}}- \frac{3}{n} \left\{
%2 H^2 + \frac{ke^{-2Ht}}{a_0^2} \right\}\right],~w =- 1,~H={\mbox constant}~.\label{rew4}
%
%\end{eqnarray}

Since we are primarily interested in the corrected Raychaudhuri equation close to the putative Big-Bang
singularity, i.e. pre-inflation with $w\neq -1$, we ignore the energy-momentum
term in Eqs.(\ref{rew2},\ref{rew3}), since this term becomes sub dominant at
larger curvatures.
Furthermore, omitting exotic matter from our discussions, i.e. $w\nleq -1/3$ such that the $k$ term can also
be ignored, we arrive at the rather simple form for the Raychaudhuri equations
\begin{eqnarray}\label{rew5}
\frac{d\hat\theta}{d\lambda} && = -\frac{\hat\theta^2}{2} + \frac{2(n-1)(2n-1) (n^t)^2}{n\, t^2}\\
\frac{d\theta}{d\tau} && = -\frac{\theta^2}{3} + \frac{2(1-3w_{\rm eff})}{3n(1+w_{\rm eff})^2t^2}~\label{rew6}
= -\frac{\theta^2}{3} + \frac{3(4n^2-5)(2n^2-3n+1)}{(n-2)^2 t^2}
\end{eqnarray}
As can be seen, the second term in both Eqs.(\ref{rew5},\ref{rew6}) are repulsive and dominate over the first term 
(due to the factor of $1/t^2$ therein) for all $n$.
This causes geodesics to not to converge in this epoch, potentially preventing the formation of a singularity. 
At later epochs, e.g. during inflation and thereafter, that term is sub-dominant as are additional terms 
originating in an actual perfect fluid described by the equation $p=w\rho$, with $-1 \leq w \leq 1/3 $.
Therefore the standard conclusions from the Raychaudhuri equation 
and the singularity theorems hold. 
Note that while in \cite{Albareti:2012va},
the authors consider constant curvature cosmological
metrics (Einstein spaces), in \cite{Santos:2016vjg} the authors consider theories of the form
$f(R) = R + \alpha/R^n$ with $\alpha <0$ and $n \in \mathbb{R}$. These are relevant to late 
time acceleration, as an alternative to dark energy. As mentioned above, here we are
primarily concerned with very early times, to examine effects of repulsive terms (if any) near the 
initial singularity. Furthermore, similar to the above references we too assume that the standard energy conditions are valid for the cosmological fluids.

\section{Loop Quantum Cosmology}
\label{sec:Loop}

\noindent

In loop quantum cosmology, one starts with the standard large scale homogeneous and isotropic model of the Universe, 
described by metric (\ref{frw2}), and obtains an effective Hamiltonian, incorporating the discrete quantum nature
of spacetime at the fundamental level, quantum back-reaction and the behavior of the scale factor at very small length scales
\cite{Ashtekar:2008ay,Singh:2009mz,Li:2008tc}
\begin{eqnarray}
{\cal H}_{\rm eff} = - \frac{6}{\kappa_4^2 \gamma^2}
\frac{\sin^2 (\lambda\beta)}{\lambda^2} V + {\cal H}_{\rm matter}~, 
\end{eqnarray}
where $\lambda$ is a measure of fundamental discreteness, $\gamma$ the Immirizi parameter, $V$ the volume, 
$\beta (=\gamma H)$ its conjugate, and ${\cal H}_{\rm matter}$ the matter Lagrangian.
Next, using the Hamiltonian constraint ${\cal H}_{\rm eff} \approx 0$ and the Hamilton's equation for $V$,
namely $\dot V = \{V, {\cal H}_{eff} \}$, one obtains the `{\it Loop Quantum Corrected Raychaudhuri Equation}'
\begin{eqnarray}
&& \frac{\ddot a}{a} = \dot H + H^2 = 
 - \frac{\kappa_4^2}{4} \left[  
\rho \left( 1 - \frac{\rho}{\rho_c} \right)
+ 3 \left\{ P \left(1-\frac{2\rho}{\rho_c} \right) - \frac{\rho^2}{\rho_c} 
\right\}
\right]\, ,
\end{eqnarray}
where: $\rho_c \equiv \frac{3}{8\pi \gamma^2 \lambda^2 V^2}$.

By substituting \cite{AKR,Haque:2017nln}
\begin{eqnarray}
\frac{\dot a}{a} &=& \frac{\theta}{3}~,  \\
\hat\theta &=& \frac{2}{a^2} \left[ \dot a - \frac{1}{r} \right]
\end{eqnarray}
for timelike and null geodesics, respectively, we can
obtain the corresponding Raychaudhuri equations
\begin{align}
\frac{d\theta}{d\tau} &= - \frac{\theta^2}{3} 
 - \frac{\kappa_4^2}{4} \left[  
\rho \left( 1 - \frac{\rho}{\rho_c} \right)
+ 3 \left\{ P \left(1-\frac{2\rho}{\rho_c} \right) - \frac{\rho^2}{\rho_c} 
\right\}
\right] ~&\mbox{(timelike)}, \label{lqcre3}\\
\frac{d\hat{\theta}}{d\lambda} &= 
- \frac{\hat\theta^2}{2} + \frac{2}{a^2} \dot H &\nonumber \\
%- \theta^2 a - \frac{2\theta}{ar} + \frac{2\dot r}{a^2 r^2} 
~~~~~ &= {\hat\theta^2} - \frac{2}{(a^2 r)^2} - \frac{2\theta}{a^2 r}   
- \frac{\kappa_4^2}{2a^2} 
\left[ \rho \left( 1 - \frac{\rho}{\rho_c} \right)
+ 3 \left\{ P \left(1-\frac{2\rho}{\rho_c} \right) - \frac{\rho^2}{\rho_c} 
\right\}
\right] 
%\dot \theta = - \frac{\theta^2}{2} 
% - {2\pi G} \left[  
%\rho \left( 1 - \frac{\rho}{\rho_c} \right)
%+ 3 \left\{ P \left(1-\frac{2\rho}{\rho_c} \right) - \frac{\rho^2}{\rho_c} 
%\right\}
%\right] 
~&\mbox{(null)}. 
\label{lqcre4}
\end{align}
Note that there are three positive (repulsive) terms in the RHS of the RE (\ref{lqcre3}), (\ref{lqcre4}),
effective near $\rho \simeq \rho_c$, i.e. in the early Universe. 
Since the above equations were derived using a {\it non-perturbative} quantization scheme, without assuming a fixed background metric
and allowing for back-reaction, one expects that they will hold good all the way to $t\rightarrow 0$. At this point, the repulsive terms
take over (since $\rho \rightarrow \infty$) and the singularity theorems and their conclusions cease to hold.

\section{Conclusions}
\label{sec:conclusion}

%\tred{SD-check $f(R)$ and LQC sections.}\\
In this paper, we have examined the effects of several different alternative theories of gravity on the geodesic convergence properties of the Raychaudhuri equation
in some simple backgrounds, and found some cases where these corrections provide repulsive terms.

In the case of string theory, we studied the leading order corrections to the Einstein Equations (for constant dilaton and vanishing form fields)
arising in the form of Einstein-Gauss-Bonnet gravity. We did not find repulsive contributions to the Raychaudhuri equation for
$D$-dimensional black hole backgrounds, which is consistent with known exact black hole solutions in Einstein-Gauss-Bonnet gravity, which still contain
singularities. We did find repulsive terms for $D$-dimensional cosmological backgrounds, however these terms do not apear to be significant enough
to prevent the existence of a Big Bang singularity, at least at leading order.
It is therefore interesting that the string theory corrections to the Raychaudhuri equation do not have a definite sign, but instead the sign of the corrections
depends on the background under consideration.
We note that our results here should be taken as preliminary steps towards analyzing the potential for string theory corrections to resolve
singularities. In particular, we have only examined the leading-order pure gravity corrections, setting the dilaton to a constant and form fields
to zero. Allowing these fields to be dynamical will give rise to additional terms in the Raychaudhuri equation that could act repulsively.
Additionally, the leading corrections we considered here vanish for type IIA/IIB string theory; it would be interesting,
though technically challenging, to include the next order of corrections.

We also considered corrections to the Einstein equations coming from induced gravity on a brane embedded in a warped 5-dimensional bulk.
In this case, the corrections show up as quadratic in the brane-localized energy-momentum tensor. For a cosmological background we found that
the corrections to the null Raychaudhuri equation always increase convergence, while the corrections ot the timelike Raychaudhuri equation
can give rise to repulsive terms when the brane matter is described by a perfect fluid with pressure $p < -2/3\ \rho$.
Nonetheless, it appears that a Big Bang singularity persists in this case as well. It would be interesting
to study these corrections for black hole backgrounds, as well as explore the role that bulk fields play in providing additional
constraints.

We examined corrections for curvature-dominated 
$f(R)$ theories of the form $f(R) = R [1+(\ell_{Pl} R)^{n-1}]$, where $\ell_{Pl}$ is the Planck length and $n$ is a positive integer.
We found that the $f(R)$ corrections can contribute repulsive terms to the timelike and null Raychaudhuri equations for $w_{\rm eff} > -1/3$, where $w_{\rm eff}(n) = -(6n^2-7n-1)/(6n^2-9n+3)$ is the effective equation of state of the curvature corrections.
It would be interesting to study these corrections further to examine whether they do indeed lead to a resolution of the cosmological singularity.

Finally, we also considered corrections to the Friedmann equation for Loop Quantum Cosmology.
These corrections give rise to repulsive terms in both the timelike and null Raychaudhuri equations, suggesting a resolution of the Big Bang singularity.
%\tred
% The following added in revised version, May 2018
{Note that in this case the results are non-perturbative in nature. 
It is important to explore $f(R)$ theories and Loop Quantum Cosmology further to see if indeed all possible geodesics are
complete in these theories, and if so, whether there exists another criterion for the existence of singularities. If not, the
corresponding spacetimes should play an important role near the normally singular regions inside black holes and in cosmology. 
}

In all of our examples we have considered simple isotropic and/or homogeneous backgrounds, and it would be interesting to study how robust our results are
to anisotropic or inhomogeneous deviations.
% Added June 2018
Since the power of the singularity theorems is in their general applicability, and generic situations likely would not possess such a high degree of symmetry, it is not clear whether the additional terms to the Raychaudhuri equation would continue to be repulsive in more general backgrounds with less symmetry. We leave this general analysis for future work.
In addition, all of our analysis has focused on the corrections to the classical equations of motion from alternative theories of gravity.
It would be interesting to examine quantum effects
(which depend on the wavefunction of the fluid) over and above the above classical terms. However, since
these effects are always repulsive in nature, and 
prevents the quantal trajectories (quantum counterparts of geodesics) from crossing, such effects should reinforce the above conclusion \cite{Das:2013oda,Alsaleh:2017ozf}.

\vspace{.2cm}
%%%%%%%%%%%%%%%%%%%%%%%%%%%%%%%%%%%%%%%%%%%%%%%%%%%%%%
\noindent {\bf Acknowledgment}

\noindent
This work is supported by the Natural Sciences and Engineering
Research Council of Canada. SSH would like to thank Perimeter Institute, where part of the work was done. DJB is supported by a PHD fellowship from the South African National Institute for Theoretical Physics. NM is supported by the South African Research Chairs Initiative
of the Department of Science and Technology and the National Research Foundation of South Africa. Any opinion, finding and conclusion or recommendation expressed in this material is that of the authors and the NRF does not accept any liability in this regard. 

\appendix

\section{Raychaudhuri Equations in $D$ Dimensions}
\label{app:RayEq}

\subsection*{Timelike}
Consider a congruence of timelike geodesics parameterized
by proper time $\tau$ with tangent vectors $u^M(\tau)$ satisfying
$u^M u_M = -1$ and $u^M \nabla_M u^N = 0$
for some $D$-dimensional metric $g_{MN}$.
We can define the ``spatial'' (or transverse) metric as
\be
h_{MN} \equiv g_{MN} + u_M u_N\, ,
\ee
which is transverse to the tangent vectors $u^M h_{MN} = 0$.
We will define the following quantities:
\be
\theta \equiv \nabla_M u^M && \mbox{ Expansion scalar } \\
\sigma_{MN} = \frac{1}{2} \left(\nabla_M u_N + \nabla_N u_M\right) - \frac{1}{D-1} h_{MN} \theta && \mbox{ Shear tensor } \\
\omega_{MN} = \frac{1}{2} \left(\nabla_M u_N - \nabla_N u_M\right) && \mbox{ Twist tensor }
\ee
Note that $\sigma_{MN}, \omega_{MN}$ are purely spatial (or transverse), since
$u^M \sigma_{MN} = 0 = u^M \omega_{MN}$.
In addition, $\omega_{MN} = 0$ for tangent vectors that are hypersurface orthogonal, as will
be the case for all of the tangent vectors considered in this paper.
Following \cite{Wald}, the expansion obeys the Raychaudhuri equation
\be
\frac{d\theta}{d\tau} = -\frac{\theta^2}{D-1} - \sigma_{MN} \sigma^{MN} + \omega_{MN} \omega^{MN} - R_{MN} u^M u^N\, .
\label{eq:TimelikeRay}
\ee
Since $\sigma_{MN}$ is purely spatial, the second term is non-positive, and as mentioned above we will
be considering cases where $\omega_{MN} = 0$, so that the right-hand side of Eq.(\ref{eq:TimelikeRay})
is non-positive as long as $R_{MN} u^M u^N \geq 0$.

\subsection*{Null}
The derivation of the null Raychaudhuri equation proceeds in a similar way as the timelike case above,
with an additional complication in identifying the transverse directions to the null geodesic.

Consider a congruence of null geodesics parameterized by affine parameter $\lambda$ with
tangent vectors $n^M(\lambda)$, satisfying $n^M n_M = 0$ and $n^M \nabla_M n^N = 0$.
In order to define the transverse directions to the geodesic we need an ``auxiliary"
null vector $k^M$ such that\footnote{$k^M$ need not be a geodesic, and indeed in most cases it is not possible 
for both $k^M$ and $n^M$ to be geodesics and cross-normalized to a constant.} $k^M k_M = 0$ and $n^M k_M = -1$.
We define the transverse metric to the null geodesics
\be
\hat h_{MN} = g_{MN} + n_M k_N + k_M n_M
\ee
which has dimension $g^{MN} \hat h_{MN} = D-2$ and is transverse to both $n^M$ and $k^M$:
$n^M \hat h_{MN} = 0 = k^M \hat h_{MN}$.
We then proceed as in the timelike case by defining:
\be
\hat \theta \equiv \nabla_M n^M && \mbox{ Expansion scalar } \\
\hat \sigma_{MN} \equiv \frac{1}{2} \left(\nabla_M n_N + \nabla_N n_M \right)- \frac{1}{D-2} \hat h_{MN} \hat \theta && \mbox{ Shear tensor } \\
\hat \omega_{MN} \equiv \frac{1}{2} \left(\nabla_M n_N - \nabla_N n_M\right) && \mbox{ Twist tensor }
\ee
As before, $\sigma_{MN}, \omega_{MN}$ are transverse to the null geodesics.
The expansion obeys the Raychaudhuri equation
\be
\frac{d\hat \theta}{d\lambda} = -\frac{\hat \theta^2}{D-2} - \hat \sigma_{MN} \hat \sigma^{MN} + \hat \omega_{MN} \hat \omega^{MN} - R_{MN} n^M n^N\, .
\label{eq:NullRay}
\ee
As in the timelike case, $\omega_{MN} = 0$ for geodesics that are hypersurface orthogonal; further, since $\sigma_{MN}$ is purely
spatial the second term in Eq.(\ref{eq:NullRay}) is non-positive, so that the entire right-hand side is non-positive as well
as long as $R_{MN} n^M n^N \geq 0$.

\section{Gauss-Bonnet Corrections in Anisotropic Cosmology}

We will examine the correction terms $H_{MN}$ for a $D$-dimensional background
consisting of a $d$-dimensional flat, homogeneous spacetime, with coordinates $\{r,\theta^i\}$,
and a $m$-dimensional homogeneous spacetime, with coordinates $\{y^m\}$, each with their own
scale factor
\begin{equation}
ds^2 = -dt^2 + a(t)^2 \left(dr^2+ r^2 \hat{d\Omega}_{d-1}^2\right) + b(t)^2 \tilde g_{mn} dy^m dy^n\,,
\end{equation}
where $D = d + m + 1$.

The Gauss-Bonnet scalar is
\be
R_{GB}^2 &=& R_{ABCD}R^{ABCD} - 4R_{MN} R^{MN} + (R_D)^2 \nonumber \\
	&=& 4d(d-1)(d-2)\frac{\ddot a}{a} \frac{\dot a^2}{a^2} + d(d-1)(d-2)(d-3)\frac{\dot a^4}{a^4} \nonumber \\
	&& + 4 m(m-1)(m-2) \frac{\ddot b}{b}\frac{\dot b^2}{b^2} + m(m-1)(m-2)(m-3) \frac{\dot b^4}{b^4} \nonumber \\
	&& + 4 dm(m-1) \frac{\ddot a}{a} \frac{\dot b^2}{b^2} + 4 dm(d-1) \frac{\dot a^2}{a^2}\frac{\ddot b}{b}
		+ 2 dm(d-1)(m-1) \frac{\dot a^2}{a^2} \frac{\dot b^2}{b^2} \nonumber \\
	&& + \frac{\t R_m}{b^2} \left[4 d \frac{\ddot a}{a} + 2 d(d-1) \frac{\dot a^2}{a^2} + 4 (m-2) \frac{\ddot b}{b^3}
		-6m(m-1) \frac{\dot b^2}{b^2} - 3 \frac{\t R_m}{b^2}\right]\,.
\ee
Note that $R_{GB}^2$ simplifies considerably when $d=3$ and $m=1,2$,
corresponding to $D=5$ or $6$-dimensional spacetime, respectively, and the internal space is flat $\t R_m = 0$
\be
R_{GB}^2|_{d=3, m=1,2} = 24\frac{\ddot a \dot a^2}{a^3} + 24 m \frac{\dot a^2}{a^2} \frac{\ddot b}{b}\, ,
\ee
though this doesn't seem to give any particularly useful interpretation.

We can now compute the correction terms
\be
H_{tt} &=& \frac{g_{tt}}{2} R_{GB}^2 - 2 R_D R_{tt} + 4R_{tt} g^{tt} R_{tt} + 4 (g^{rr})^2 R_{rr} R_{trtr} + 4 g^{ii'}g^{jj'} R_{i'j'} R_{titj} + 4 R^{mn} R_{tmtn} - 2  R_{tABC} R_t^{ABC} \nonumber \\
	&=& -\frac{1}{2} d(d-1)(d-2)(d-3) \frac{\dot a^4}{a^4} - \frac{1}{2} m(m-1)(m-2)(m-3) \frac{\dot b^4}{b^4} - d m (d-1)(m-1) \frac{\dot a^2}{a^2} \frac{\dot b^2}{b^2}\nonumber \\
		&& + \frac{\t R_m}{b^2} \left[-d(d-1) \frac{\dot a^2}{a^2} + 3 m(m-1) \frac{\dot b^2}{b^2} + \frac{3}{2} \frac{\t R_m}{b^2}\right]\,; \\
H_{rr} &=& \frac{g_{rr}}{2} R_{GB}^2 - 2 R_D R_{rr} + 4 R_{rr} g^{rr} R_{rr} +4 R_{tt} (g^{tt})^2 R_{rtrt} + 4 R_{i'j'}g^{ii'}g^{jj'} R_{rirj}- 2 R_{rABC}R_r^{ABC}  \nonumber \\
	&=&  2(d-1)(d-2)(d-3) \frac{\ddot a \dot a^2}{a} + \frac{1}{2} (d-1)(d-2)(d-3)(d-4)\frac{\dot a^4}{a^2} \nonumber \\
		&& + 2 m(m-1)(m-2) a^2 \frac{\ddot b \dot b^2}{b^3}+ m(m-1)(m-2)(m-3) a^2 \frac{\dot b^4}{b^4} \nonumber \\
		&& + 2 m(m-1)(d-1) a \ddot a \frac{\dot b^2}{b^2} + 2 m(d-1)(d-2) \dot a^2 \frac{\ddot b}{b} + m(m-1)(d-1)(d-2) \dot a^2 \frac{\dot b^2}{b^2} \nonumber\\
		&& + a^2 \frac{\t R_m}{b^2} \left[2d\frac{\ddot a}{a} + d(d-1) \frac{\dot a^2}{a^2} + 2(m-2) \frac{\ddot b}{b^3} - 3 m(m-1) \frac{\dot b^2}{b^2} - \frac{3}{2} \frac{\t R_m}{b^2}\right]\,.
\ee
Notice that $H_{tt}$ vanishes identically for all $d =3$, $m=1,2$ and a flat internal space $\t R_m = 0$, 
while $H_{rr}$ is considerably more complex.

\subsubsection{Raychaudhuri Corrections}

We now consider a $d$-dimensional radial, null, affine tangent vector
\be
n^M = \left(\frac{1}{a(t)},\pm \frac{1}{a(t)^2},\vec{0},\vec{0}\right)\, .
\ee
Note that $n^M n_M = 0$ (it is null) and $n^M$ satisfies the affine condition $n^N \nabla_N n^M = 0$, even for the anisotropic background considered here.

The null Raychaudhuri equation takes the form
\be
\frac{d\hat \theta}{d\lambda} = \frac{-\hat \theta^2}{D-2}-|\hat \sigma|^2 - R_{MN}n^M n^N\, .
\ee
Our corrections due to the Gauss-Bonnet term appear on the right hand side, as
\be
&&\frac{8}{\alpha'}R_{MN}n^M n^N \sim   H_{MN}n^M n^N = H_{tt} n^t n^t + H_{rr} n^r n^r \nonumber \\
	&&= \frac{(d-1)(d-2)(d-3)(d-5)}{2} \frac{\dot a^2}{a^6} + 2 (d-1)(d-2)(d-3) \frac{\ddot a \dot a^2}{a^5} \nonumber \\
		&& + \frac{(m-1)(m-2)(m-3)(m-5)}{2} \frac{\dot b^4}{a^2 b^4} + 2 m(m-1)(m-2) \frac{\ddot b\dot b^2}{a^2 b^3} + 2 m (d-1)(d-2) \frac{\dot a^2}{a^4} \frac{\ddot b}{b} \nonumber \\
		&& + 2 m(m-1)(d-1) \frac{\dot b^2}{b^2} \left[\frac{\ddot a}{a^3}-\frac{\dot a^2}{a^4}\right] + 2 (m-1) \t R_m \frac{\ddot b}{a^2 b^5}\, .
\ee
As  before, choosing $d =3$, $m=1,2$ and a flat internal space $\t R_m = 0$ simplifies the result considerably
\be
H_{MN}n^M n^N &=& 2m(m-1)(d-1) \frac{\dot b^2}{b^2}\left[\frac{\ddot a}{a^3} -\frac{\dot a^2}{a^4}\right] + 2 m(d-1)(d-2) \frac{\dot a^2}{a^4} \frac{\ddot b}{b} \nonumber \\
 &=& 4m(m-1) \frac{\dot b^2}{b^2} \frac{\dot H_d}{a^2} + 4 m\frac{\dot a^2}{a^4} \frac{\ddot b}{b}\, ,
\ee
where we wrote $H_d \equiv \dot a/a$, and thus $\dot H_d = \ddot a/a - \dot a^2/a^2$.
The overall sign of this correction term is not clear; typically we expect $\dot H_d \leq 0$ for spacetimes that don't have
NEC violating matter, so this term is potentially negative. However, it is unclear the scenarios for which $\ddot b/b < 0$.

Finally, consider a comoving, proper-time parameterized timelike geodesic described by the tangent vector
\be
u^M = \left(1,0,\vec{0},\vec{0}\right)\,.
\ee
It is straightforward to check that $u^M u_M = -1$ and $u^N \nabla_N u^M = 0$.

The timelike Raychaudhuri equation takes the form
\be
\frac{d\theta}{d\tau} = -\frac{\theta^2}{D-1} - |\sigma|^2 - R_{MN} u^M u^N\, ,
\ee
where again the corrections to the Raychaudhuri equation due to the Gauss-Bonnet corrections come from the last term
\be
&&\frac{8}{\alpha'} R_{MN} u^M u^N \sim  H_{MN} u^M u^N = H_{tt} u^t u^t \nonumber \\
	&&= -\frac{d(d-1)(d-2)(d-3)}{2} \frac{\dot a^4}{a^4} - \frac{m(m-1)(m-2)(m-3)}{2} \frac{\dot b^4}{b^4} -dm(d-1)(m-1) \frac{\dot a^2}{a^2} \frac{\dot b^2}{b^2}\nonumber \\
	&& + \frac{\t R_m}{b^2}\left[-d(d-1)\frac{\dot a^2}{a^2}+3m(m-1) \frac{\dot b^2}{b^2} + \frac{3}{2} \frac{\t R_m}{b^2}\right]\, .
\ee
Notice here that for a flat internal space $\t R_m = 0$, these corrections are {\bf manifestly negative}, thus they cause
divergence in the Raychaudhuri equation.

More generally, we see that whether these additional correction terms due to Gauss-Bonnet act 
to increase or oppose convergence depends on the particular background, and whether one is considering timelike or
null rays. Thus, there are backgrounds in which the usual singularity theorems no longer hold.

%%%%%%%%%%%%%%%%%%%%%%%%%%%%%%%%%%%%%%%%% SHAJID%%%%%%%%%%%%%%%
%%%%%%%%%%%%%%%%%%%%%%%%%%%%%%%%%%%%%%%%%
\section{Weyl Curvature Tensor}
\label{sec:Weyl}

In Section \ref{sec:braneworld}, we encountered
corrections to the induced 4-dimensional
Einstein equations that involved the 5-dimensional
Weyl curvature through the tensor
\be
E_{\mu\nu} &=& ^{(5)}C_{\mu\alpha\nu\beta} n^\alpha n^\beta\, .
\label{eq:EmunuAppendix}
\ee

In this Appendix, we collect some useful properties
of the Weyl tensor, as well as evaluate this term for the braneworld metric (\ref{eq:BraneMetric}).

\subsection{Weyl Tensor}
We begin with some facts about the Weyl curvature tensor. First, the definition of the Weyl curvature tensor is
\be
C_{\alpha\beta\gamma\delta} = R_{\alpha\beta\gamma\delta} + \frac{1}{n-2} \left[g_{\alpha\delta} R_{\gamma \beta} + g_{\beta\gamma} R_{\delta \alpha} - g_{\alpha \gamma} R_{\delta \beta} - g_{\beta\delta} R_{\gamma \alpha}\right]
	+ \frac{1}{(n-1)(n-2)} \left[g_{\alpha \gamma} g_{\delta\beta} - g_{\alpha\delta} g_{\gamma\beta}\right] R
\ee
in terms of the Riemann tensor, Ricci tensor, and Ricci scalar. In effect, the Weyl tensor $C_{\alpha\beta\gamma\delta}$ is the traceless part of the Riemann curvature tensor.
As such, it inherits the usual symmetry identities from the Riemann tensor, as well as a traceless condition:
\be
&&C_{\alpha \beta \gamma \delta} = - C_{\alpha \beta \delta \gamma} = -C_{\beta\alpha \gamma \delta} = C_{\gamma \delta \alpha \beta} \\
&& C^\alpha_{\beta \alpha \delta} = 0 \hspace{.2in} \mbox{(true for any 2 contracted indicies)}
\ee
The Weyl curvature tensor has an interesting feature: for any 2 metrics that can be related by a conformal factor
\be
\hat g_{\alpha \beta} = \Omega^2 g_{\alpha \beta}\, ,
\ee
the Weyl curvature tensors are {\bf the same}
\be
\hat C_{\alpha \beta\gamma\delta} = C_{\alpha \beta\gamma \delta}\, .
\ee
This means that if a metric can be written in a conformally flat form, then $C_{\alpha\beta\gamma\delta} = 0$ identically.

In particular, for the RS Minkowksi metric (\ref{eq:RSmetric}), this can be written in conformally flat form (with the coordinate redefinition of $d\chi = e^{-K(z)} dz$).
Further, the ansatz metric (\ref{eq:BraneMetric}), is {\bf almost} conformally flat, aside from the factor $b(t,y)$; thus, we expect
the Weyl curvature tensor for (\ref{eq:BraneMetric}) to be proportional to the time- and space-derivatives of $b(t,y)$ (as we subsequently find it to be).
%%%%%%%%%%%%%%%%%%%%%%%%%%%%SHAJID%%%%%%%%%%%%%%%%%%%%%
\subsection{Evaluating $E_{\mu\nu}$}

In Section \ref{sec:braneworld} we evaluated the 
$E_{\mu\nu}$ tensor 
in the limit $y\rightarrow 0$, 
finding that $E_{\mu\nu} \propto \mu \frac{b^2}{a^2}$, consequently the singular terms do not contribute.
However, we need to examine the consistency of this result.
In particular, the induced Einstein equations (\ref{eq:inducedEE}) are not closed, and must be supplemented
by additional constraints \cite{Shiromizu:1999wj}.
The Bianchi identity for the induced Einstein equations (\ref{eq:inducedEE}) implies (assuming that the brane EM tensor $\tau_{\mu\nu}$ obeys EM conservation (\ref{eq:EMConservation}))
\be
\hat \nabla^\mu E_{\mu\nu} = \kappa_5^2 \hat\nabla^\mu \pi_{\mu \nu}\, ,
\label{eq:Bianchi}
\ee
where $\hat \nabla^\mu$ is the covariant derivative constructed with respect to the 4d $\hat q_{\mu\nu}$ induced metric.
Thus, we will examine the consistency of our limiting procedure by verifying that our result for $E_{\mu\nu}$ is indeed a solution of the Bianchi identity (\ref{eq:Bianchi}).

Starting on the right-hand side, we rewrite the Bianchi identity as $\hat \nabla^\mu \pi_{\mu\nu} = \hat \nabla_\mu \pi^\mu_\nu$, and using the suitably raised versions of the $\pi^\mu_\nu$ tensor,
it is straightforward to compute
\be
\hat \nabla^\mu \pi_{\mu t} = -\frac{1}{6} \rho \dot \rho - \frac{1}{2}\frac{\dot a}{a} \rho (p+\rho) = -\frac{\rho}{6}\left(\dot \rho + 3\frac{\dot a}{a}(\rho+p)\right) = 0\, ,
\ee
where we used energy conservation of the brane matter (\ref{eq:EMConservation}) to set this entire term to zero.
Similar results are obtained for the $\nu = r,\theta,\phi$ components of the Bianchi identity (\ref{eq:Bianchi}).

However, for the left-hand side of the $\nu = t$ component of the Bianchi identity, we obtain
\be
\hat \nabla_\mu E^\mu_t = 3\frac{\dot a}{a} \frac{\mu}{a^4} (1-b^2)\, .
\ee
The only way to satisfy the Bianchi identity (\ref{eq:Bianchi}), then, is to require the dark radiation term to vanish, $\mu = 0$.
Similar results are obtained for the $\nu = r, \theta, \phi$ components.

%%%%%%%%%%%%%%%%%%%%%%Brane Appendix end%%%%%%%%%
%%%%%%%%%%%%%%%%%%%%%%

\bibliographystyle{utphysmodb}

\bibliography{refs}

\end{document}